# Mechanisms of Molecular Ferroelectrics Made Simple

*Xiaoqing* Zhu†,‡, Wenbin Fan†,‡, Wei Ren†,‡, and Yongle Li*,†,‡

† Department of Physics, Shanghai University, Shanghai 200444, China.

‡ Department of Physics, International Center of Quantum and Molecular Structures, Shanghai University, Shanghai 200444, China.

*Email: yongleli@shu.edu.cn;

Molecular ferroelectrics have captured immense attention due to their superiority over inorganic oxide ferroelectrics, such as environmentally friendliness,, low cost, flexibility, and foldability. However, the mechanisms of ferroelectric switching and phase transition for the molecular ferroelectrics are still missing, which leads to the development of less efficient novel molecular ferroelectrics. In this work, we have provided a protocol by combining molecular dynamics simulation on a polarized force field named polarized crystal charge and an enhanced sampling technique, replica-exchange molecular dynamics, to simulate such mechanisms. With this procedure, we have investigated a promising molecular ferroelectric material, ($R$)/($S$)-3-quinuclidinol crystal. We have simulated the ferroelectric hysteresis loops of both enantiomers and obtained spontaneous polarization as 7.4±0.1 and 7.8 ± 0.1 $\mu C \cdot cm^{-2}$, respectively, and the corresponding coercive electric field as 15 $kV \cdot cm^{-1}$. We have also found the Curie temperature ($T_c$) to be 380/385 K for ferro-/paraelectric phase transition of both enantiomers. All of the simulated results are highly compatible with experimental values. In addition to this, we predict a novel $T_c$ of about 600 K. This finding is further validated by principal component analysis. Our work would promote the future exploration of multifunctional molecular ferroelectrics for the next generation of intelligent devices.

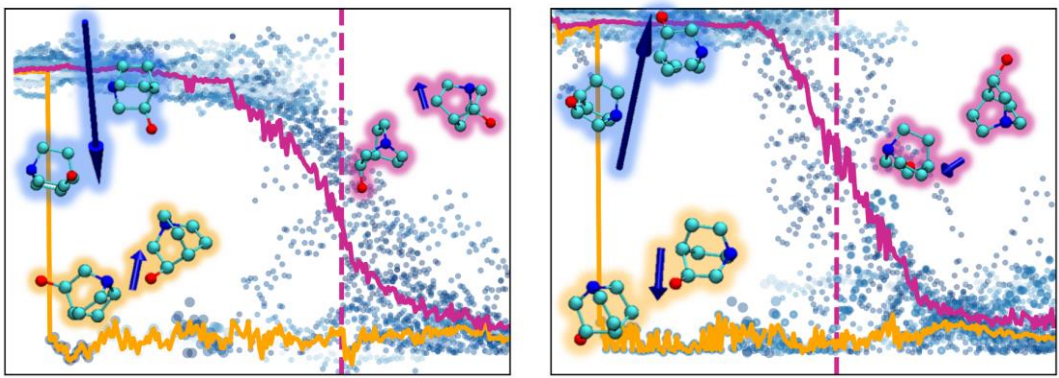

## Introduction

As a functional material in electronic technology, ferroelectric materials have the characteristics such as non-volatile ferroelectric switching effects, piezoelectricity, pyroelectricity and nonlinear optical effects, and can be used to make non-volatile ferroelectric random access memory (FeRAM)[1], miniature piezoelectric drive motors, pyroelectric detectors, light modulators, and myriads of multifunctional devices[2, 3], which are widely used in high-tech industries such as electronic communications, biomedicine, national defense technology, and aerospace.[4] Since the first ferroelectrics, Rochelle salt ($KNaC_4H_4O_6 \cdot 4H_2O$), discovered in 1920,[5] ferroelectricity is under extensive investigation until now. In the course of the development of ferroelectrics in the past century, a large number of different types of ferroelectric materials have been found, including inorganic oxides,[6, 7] molecular crystals,[8] liquid crystals,[9] polymers,[10] etc. Traditionally, the most widely applied ferroelectrics in the industry are inorganic ceramics, such as $BaTiO_3$, $Pb(Zr, Ti)O_3$ (PZT), and $Pb(Mg_{1/3}Nb_{2/3})O_3$-$PbTiO_3$ (PMN-PT), or ferroelectric polymers such as polyvinylidene fluoride (PVDF). However, due to the preparation process, inorganic ferroelectrics have disadvantages of high toxicity, high sintering temperature, difficulty in preparation, and high cost, which limit their wider application.[11] On the other hand, ferroelectric polymers are soft and flexible and can be easily customized to meet the requirements of next-generation flexible and wearable devices, though such polymers still have disadvantages, such as large coercive field ($E_c$) and uniaxial characteristicand their application is also limited.

In the recent two decades, organic molecular ferroelectric materials have become a good alternative to next-generation ferroelectrics for industrial usage, because of their multi-facet advantages, such as they are environmentally friendly, non-toxic, low cost, flexible and wearable.[12] Molecular ferroelectrics not only have the properties of high Curie temperature ($T_c$) and high saturation polarization as large as those of inorganic ferroelectrics but also have the advantages of flexibility, structural tunability, and versatility, allowing them to be utilized in a myriad of cases.[13, 14]

In 2019, a pair of molecular ferroelectric with chirality and high $T_c$, $(R)/(S)$-3-quinuclidinol, was prepared, [15] which is highly promising for industrial usage, possessing large saturation polarization ($P_s$), relatively small $E_c$, and high $T_c$. Before this, the ferroelectrics containing a single chiral molecule were limited to chiral tartaric acid and with less value for industrial applications. [16] $(R)/(S)$-3-quinuclidinol is found in chemistry and medicine applications, and its ferroelectricity was predicted by molecular design. [12] This is a breakthrough in the discovery of molecular ferroelectrics since other candidates proposed before suffer from low $T_c$. [14]

In addition to the triumphs of preparing molecular ferroelectric crystals in the study of molecular ferroelectrics, the mechanisms of the origin of spontaneous polarization, ferroelectric switching , and phase transition are still vague, and there is a lack of systematical investigations on such topic. Due to the missing of a thorough understanding of such mechanisms, the development of novel molecular ferroelectrics is still highly dependent on experience. [17] Even worse, these mechanisms also cannot all be obtained from first principles calculations. However, the confirmation of such mechanisms is a key question that needs to be solved in the investigation of molecular ferroelectrics. [18] So far, the research on molecular ferroelectricity is still far from enough, and a protocol for systematical investigation on such a topic is urgent.

Recently, our group proposed a protocol of molecular dynamics (MD) simulation for a kind of molecular ferroelectrics, the bio-ferroelectrics. [19] It can give not only reasonable predictions of $P_s$, $E_c$, and $T_c$, but also mechanisms of both ferroelectric switching and ferro-/paraelectric phase transition. This is an important basis for the study of the ferroelectric properties of materials.

For better understanding and making practical usages of the molecular ferroelectrics, and also validating our protocol, this work will study the $(R)/(S)$-3-quinuclidinol on the atomic scale, using the MD simulation methods with polarized crystal charge (PCC) to systematically elucidate the dynamics of ferroelectric switching under an external electrostatic field and phase transition with heating to obtain the mechanisms for both ferroelectric switching and

ferro-/paraelectric phase transition processes and shed a light on the *de novo* design of such kind of molecular ferroelectrics in the future. The work in this paper can not only provide a reliable protocol for future theoretical and simulation work but also verify and guide the design of molecular crystal ferroelectric materials to promote the application of organic chiral molecular crystal ferroelectric materials in production and life.

**Computational Details**

The original molecular structures were obtained from the CCDC database, [20] with entry ID 1869376 (*R*)- and 1869377 (*S*)-: (*R*)-3-quinuclidinol [ferroelectric (paraelectric) space group: [$P6_1$ ($P6_122$)] and (*S*)-3-quinuclidinol [$P6_5$ ($P6_522$)]]. We optimized the crystal structure using the first principles calculations. As reported by experimental work, the single molecule of the chiral structure of (*R*)/(*S*)-3-quinuclidinol is in point group 6 ($C_6$). [15] The first principles calculations were performed by the Vienna *Ab initio* Simulation Package (VASP)[21] based on the projector augmented wave pseudopotentials. [22] We used two exchange-correlation functionals, the pure generalized gradient approximation functional Perdew Burke Ernzerhof, [23] and hybrid functional Heyd-Scuseria-Ernzerhof (HSE). [24] The former is the most widely adopted, and the latter can give the accurate geometry and band gap for molecular crystals. Plane wave basis sets were used and the cutoff energy was set as 400 eV. The Brillouin zone was sampled with $5 \times 5 \times 1$ k-point in a Monkhorst-Pack grid. To optimize the structure, both the lattice constants and positions of all atoms are relaxed until the force is less than 0.01 eV Å$^{-1}$ for all calculations, and the criterion for the total energy was set to be $10^{-6}$ eV. The VESTA package[25] was used to visualize crystal structures. Various functionals were carried out to optimize the structure of (*R*)-3-quinuclidinol and (*S*)-3-quinuclidinol, which shows that the crystal structure calculated by the HSE functional is closer to the experimental structure compared with other functionals, while the PBE functional underestimates the band gap by about 1 - 2 eV, therefore, the HSE functional is adopted in all following calculations.

Especially, to deal with the periodic crystal model correctly, the PCC[19, 26] was incorporated into the force field recently developed by our group. In our previous work, we found that the $P_s$ of γ-glycine simulated by PCC MD was about 53 μC·cm$^{-2}$, comparing with the first principles calculation result 70.9 μC·cm$^{-2}$, with a relative error of 25%. [19] In our another work on the polarization of DIPAC, DIPAB and DIPAI, the $P_s$ from PCC MD was 5.4 ± 0.3, 5.0 ± 0.4 and 4.0 ± 0.4 μC·cm$^{-2}$ respectively, and the corresponding results from first-principles calculation is 6.8, 6.2 and 5.3 μC·cm$^{-2}$. The relative error is 21, 19 and 25%.[26] Therefore, we can conclude that our PCCcan give reliable spontaneously with moderate accuracy. Figures S3 and S4 and Tables S1 and S2 show the convergence of PCC values for all atoms. The Visual Molecular Dynamics (VMD)[27] package is used for visualization.

The MD simulations at different temperatures and under external electric fields were carried out in the NAMD software package. [28, 29] To study polarization switching, we ran several 2 ns independent trajectories with different external electric fields as shown in Figure 2 at 310 K in the NVT ensemble. In order to reduce the size problem, a 6 × 6 × 6 supercell with a size of 37.80 × 37.80 × 180.35 Å was used for (R)/(S)-3-quinuclidinol, including 1296 molecules and 28,512 atoms in total. The particle meshed ewald method was used for treating long-range electrostatic interaction[30] with a grid size of 1 Å, and the cutoff of Lennard-Jones[31] and electrostatic interaction was set to 12 Å. During the simulation process, the time step was set to 2 fs.

A random Langevin dynamics thermostat[32] and Nose–Hoover Langevin barostat were utilized for maintaining the constant temperature and pressure (NPT) ensemble, and the SHAKE algorithm[33] was used for all covalent bonds involving hydrogen atoms. The steepest descent method was used to implement energy minimization for 60,000 steps, where the maximum force value of 10 kJ/nm/mol was taken as the convergence criterion. Then, 2 ns equilibrations were performed at the NPT ensemble. The equilibration simulations were run at a constant temperature of 300 K and a constant pressure of 1 atm with 2 ns.

To obtain an equilibrium ferro-/paraelectric phase transition pathway, we performed replica-exchange molecular dynamics (REMD) simulations, [34, 35] which can more effectively sample the conformational space. The "Multisander" package in the AMBER 20[36] was used to deal with 36 replicas with a temperature range from 360 to 800 K, each trajectory is obtained from a 10 ns production run. [37] The strategy for selecting temperature is to first run a short REMD simulation in the experimental temperature and then calculate the temperature dependence of the average energy. To ensure that the system can pass the potential energy barriers, both the temperature values and the exchange frequency of temperatures of the replicas are controlled so that the acceptance rate of targets reaches 20 % and remain constant during our simulations. [35] The original structure was minimized via the first 500 cycles in the steepest descent method, and the SHAKE algorithm was utilized to constrain all the bonds with the hydrogen atoms, [38] while adopting chiral constraint. Throughout the REMD simulations, initially, replicas were heated using Langevin thermostat for 200 ps, during which copy exchanges were attempted every 10,000 steps. [39] The analysis was performed using the "ptraj" program in the Amber Tools software package. In addition, principal component analysis (PCA) was performed on the trajectory. First, the trajectories of 36 replicas were combined into a meta-trajectory for ensuring that the comparisons of all data were in the same subspace. Once the eigenvector was obtained, the meta-trajectory was projected onto the modes belonging two largest PCs. [40] PCA was performed by Cpptraj[41] in AMBER and standalone package Bio3D,[42] and all visualization analysis is performed with VMD. [43]

**Results and discussion**

The comparison of the crystal structures of ($R$)-3-quinuclidinol and ($S$)-3-quinuclidinol can be found in Figure 1. The primitive cells of crystal structures of both ($R$)- and ($S$)- 3-quinuclidinol contain six molecules separately, with a long $c$-axis ($z$-axis in the figure). Figure S1 and S2 present the *ab initio* energy band and density of states (DOS) of the($R$)/($S$)-3-quinuclidinol crystal under the HSE functional[24], which shows a gap of 5.5 eV and 5.7 eV, respectively, demonstrating the nature of the insulator. In its crystal, there are six molecules within a primitive cell, with a dipole moment of 2.45 Debye for both ($R$)- and ($S$)- enantiomers. Thus, the total dipole within the cell is 4.7 µC·cm$^{-2}$ in vacuum. The electrostatic polarization within the crystal provides about 2.0 µC·cm$^{-2}$ extra polarization, leaving a total spontaneous polarization of 6.7 µC·cm$^{-2}$.

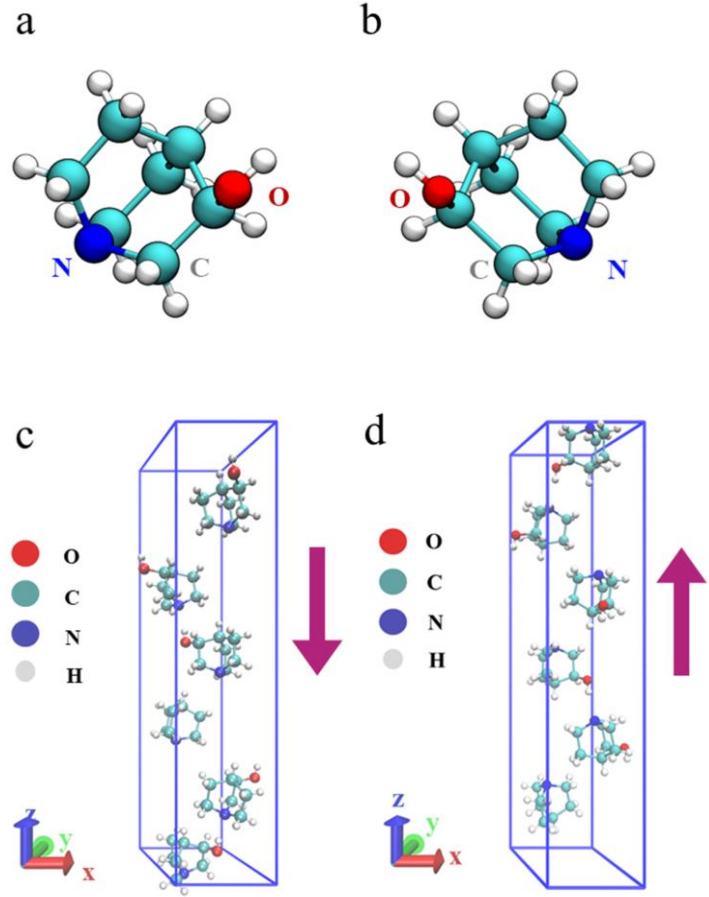

**Figure 1:** Single molecular and crystal structures of both enantiomers ($R$)-3-quinuclidinol and ($S$)-3-quinuclidinol.

a: Single molecule of ($R$)-3-quinuclidinol. b: Single molecule of ($S$)-3-quinuclidinol. c: Prime cell of ($R$)-3-quinuclidinol. d: Prime cell of ($S$)-3-quinuclidinol.

## 1. Electric Hysteresis Loop

To understand the origin of ferroelectricity of ($R$)/($S$)-3-quinuclidinol, first we examined the changing of the crystal structure and polarization under the external electric field $\boldsymbol{E}$ at 300 K. The direction of $\boldsymbol{E}$ is set along the $c$-axis, which is the $z$-direction in our model, and the magnitude of $\boldsymbol{E}$ is chosen as reported by experimental work[15]. During the MD simulation, $|\boldsymbol{E}|$ initially increased from 0 to 30 kV cm$^{-1}$, then decreased to -30 kV·cm$^{-1}$ inversely, and finally returned to 0 kV cm$^{-1}$ again.

The results are shown in Figure 2. It is clear that a fully aligned ferroelectric molecular arrangement forms after about 25 kV·cm$^{-1}$ and saturated to about 7.4 $\pm$ 0.1 μC·cm$^{-2}$ for the ($R$)- isomer and 7.8 $\pm$ 0.1 μC·cm$^{-2}$ for the ($S$)- isomer separately. and when the $\boldsymbol{E}$ reversed, the polarization switched accordingly to -10 kV·cm$^{-1}$ and then reached a reversed saturate value after $|\boldsymbol{E}|$ increased to about 25 kV·cm$^{-1}$. Such results are highly compatible with experimental results, $P_s$ = 6.69 μC·cm$^{-2}$ (for ($R$)-) and $P_s$ = 6.72 μC·cm$^{-2}$ (for ($S$)-) at 303 K. Our simulated values are also compatible to the values calculated by Xiong's group using first principles as 7 μC·cm$^{-2}$.[15] From the simulated hysteresis loop, the $E_c$ values for both ($R$)- and ($S$)- isomers are both 15 kV·cm$^{-1}$, also highly consistent with experimental results of 15 kV·cm$^{-1}$. Due to the Landauer paradox[44,45], our simulation would overestimate the $E_c$.

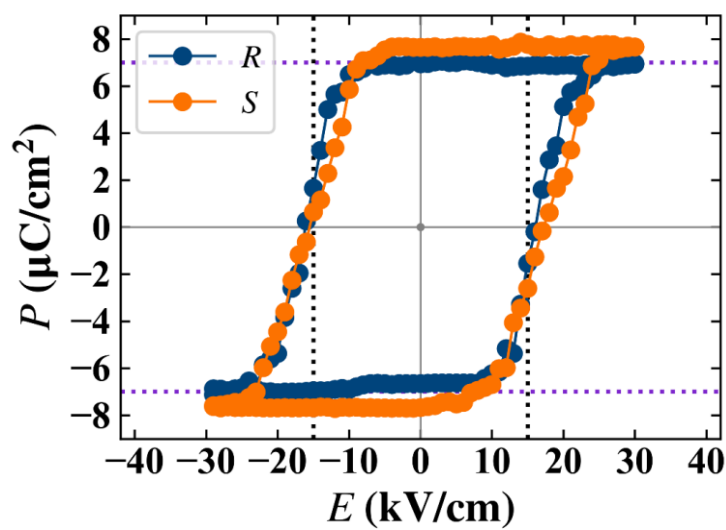

**Figure 2:** Ferroelectric hysteresis loop from MD simulation using PCC shows polarization switching of ($R$)/($S$)-3-quinuclidinol under an external electrostatic field at 300 K. From the hysteresis loop, the $P_s$ values for ($R$)- and the ($S$)- isomers are 7.4 ± 0.1 μC·cm$^{-2}$ and 7.8 ± 0.1 μC·cm$^{-2}$ respectively, and the $E_c$ value for both of the isomers is 15 kV·cm$^{-1}$.

## 2. Ferroelectric Switching Mechanism

To reveal the switching mechanism, we also plot the $P_s$ evolving with time. At the molecular level, the microscopic mechanism of the polarization switching of the $(R)/(S)$-3-quinuclidinol crystal material in the electric field is shown in Figure 3a, b. The structure consists of a $6 \times 6 \times 6$ supercell of the $(R)/(S)$-3-quinuclidinol crystal, which switches spontaneously from one orientation to the other at 300 K and keeps a constant volume under a coercive electric field of 15 kV·cm$^{-1}$ along the $c$-axis. The new domain starts nucleating at about 4.0 ns, and the ferroelectric switching process is completed after 5.50 and 5.58 ns, respectively for $(R)$- and $(S)$- isomers. The domain nuclei emerge randomly and grow quickly within about 0.2-0.3 ns, leaving most of the system switched, and then saturate gradually during another 0.3-0.5 ns. The total process occurs homogeneously in space and is lack of any pattern. but can be understood from Figure 3a and b. Since in the snapshots, it can be seen that there are multiple micro domains occurring spontaneously after about 4 ns, which violates one of the basic assumption of the Kolmogorov–Avrami–Ishibashi (KAI) model, that is, there is only a single domain existing during the whole switching process.[46]

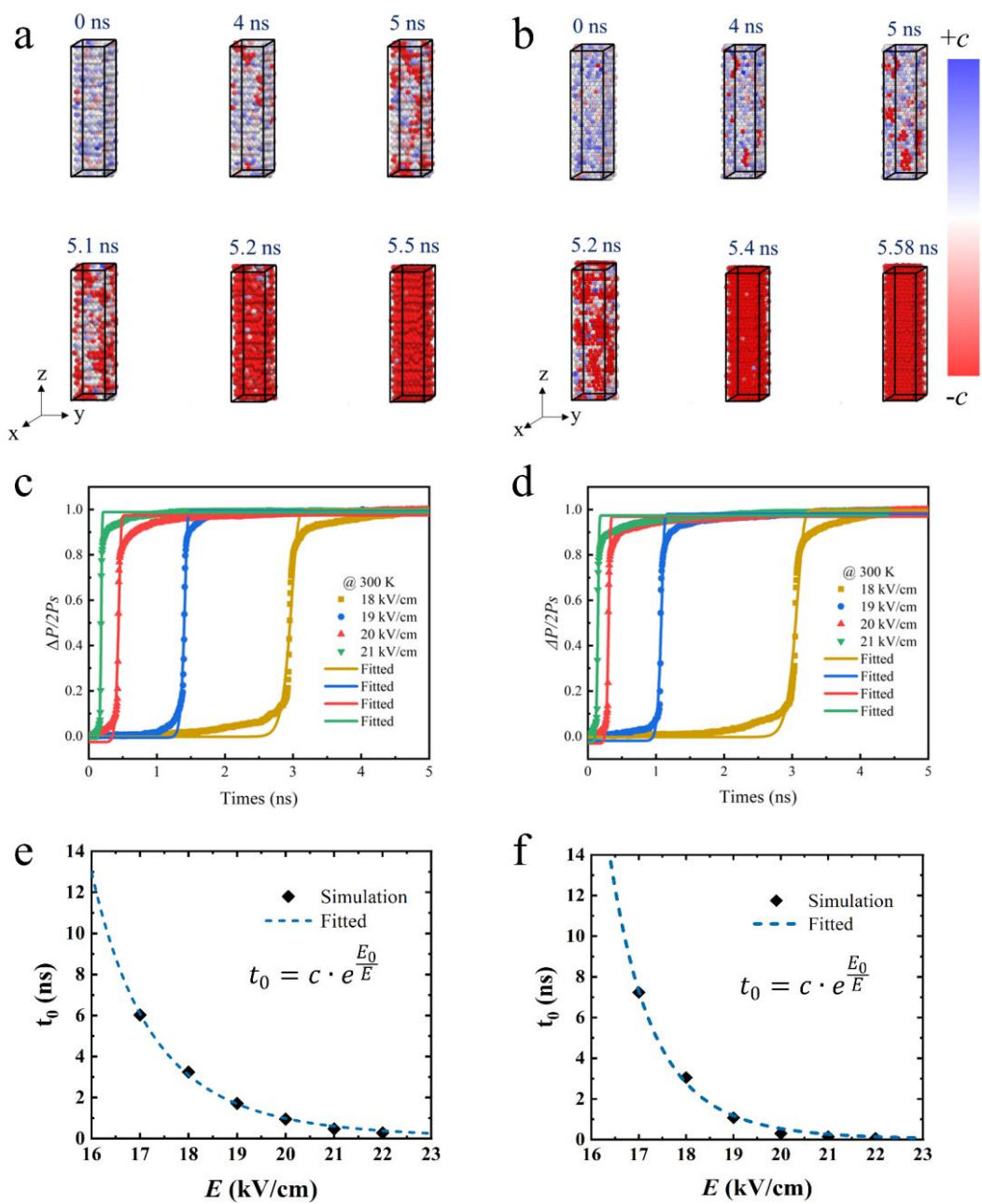

**Figure 3:** (a and b): Process of polarization switching of ($R$)/($S$)-3-quinuclidinol in an external electric field of 17 kV·cm$^{-1}$ under PCC MD simulation. The red and blue regions represent different directions of polarization, corresponding to domains with the +$c$ and -$c$ directions of their molecular dipoles. (a): ($R$)-3-quinuclidinol. (b): ($S$)-3-quinuclidinol. (c and d): The evolution of the growth of domain during the switching process under different $|E|$. The solid lines correspond to fitted results using the KAI model. (e and f): Activation electric fields with the parameter $t_0$ fitted in the KAI model. The dashed lines represent fitted functions.

For better understanding the switching mechanism, we fitted the evolution of spontaneous polarization with time to the (KAI) model, [45,47] as did in our previous work[19]:

$$\Delta P(t) = 2P_s \left( 1 - e^{-\left(\frac{t}{t_0}\right)^n} \right)$$

In the formula, the $t_0$ and $n$ are fitted parameters. The fitted lines comparing with the original data are shown in Figure 3(c-f). The converged parameters $t_0$ and $n$ are listed in Table S5. To our surprise, the switching mechanism differs much from the KAI model, especially in the time region of domain emerging and near saturating. However, in the asymptotic and switching region, the data are highly compatible with the KAI model. With the parameters from the KAI model, we also investigate the relationship between $t_0$ and the strength of the electric field $|\boldsymbol{E}|$, by fitting them to the Merz' law. [48, 49] These fittings indicate that the relationship between the characteristic conversion time of $(R)/(S)$-3-quinuclidinol and the external electric field is very consistent with that of traditional perovskite $BaTiO_3$[47, 49]:

$$t_0 = ce^{\frac{E_0}{E}}$$

The fitting activation electric fields of $(R)/(S)$-3-quinuclidinol are 100.7 kV·cm$^{-1}$ and 157.7 kV·cm$^{-1}$, respectively. Here the data show high accordance to the model, showing a universality of such ferroelectric switching.

## 3. phase Transition Mechanism

Since the practical applications of ferroelectrics need maintaining the spontaneous polarization within a specific temperature range, we investigated its phase transition under changing of temperature. First, we performed MD simulations of 他和 $(R)/(S)$-3-quinuclidinol crystal in the temperature range of 100 to 800 K, as shown in Figure 4. For $(R)$-3-quinuclidinol, all the trajectories converge within 2 ns. And when the temperature gradually gets close to 370 ± 4 K, the average value of polarization decreases with the fluctuation increasing. When the temperature reaches 375 ± 3 K, the polarization drops suddenly to 1.0 µC·cm$^{-2}$ at the very beginning indicating that the ferroelectric phase transforms into paraelectric one. Within fluctuation of MD simulation, it shows a first-order phase transition. Such finding confirms the results from second harmonic generation experiments. [15] However, this protocol fails to give reasonable $T_c$ for (*S*)-3-quinuclidinol, which shows a quite high $T_c$ as 700 K, far from the reported value of 400 K. It would stem from the fact that the free energy barrier for the phase transition of (*S*)-3-quinuclidinol is too high to allow the system to overcome it at 400 K.

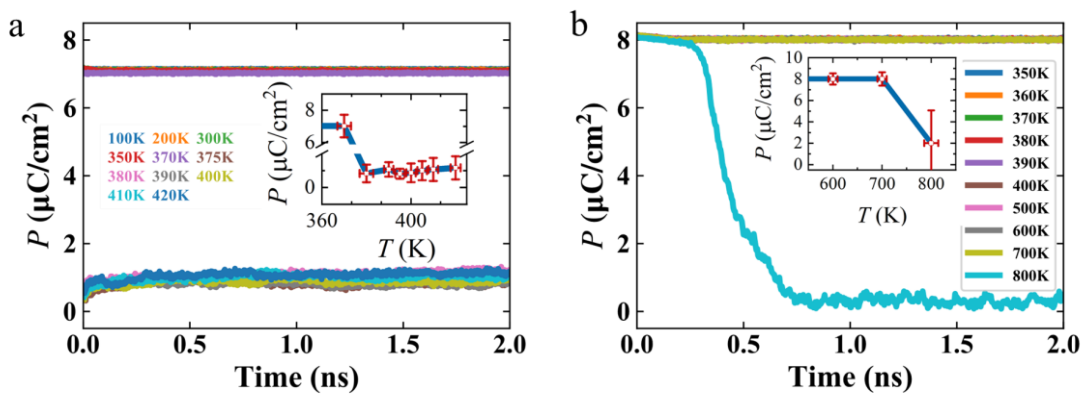

**Figure 4:** Temperature dependence curve of (a) (*R*)-3-quinuclidinol and (b) (*S*)-3-quinuclidinol. The subplot shows the polarization evolving with temperature.

To verify our consideration and obtain reliable $T_c$, we performed enhanced sampling replica-exchange REMD simulations to find the reasonable $T_c$ by effective sampling in phase space[55].

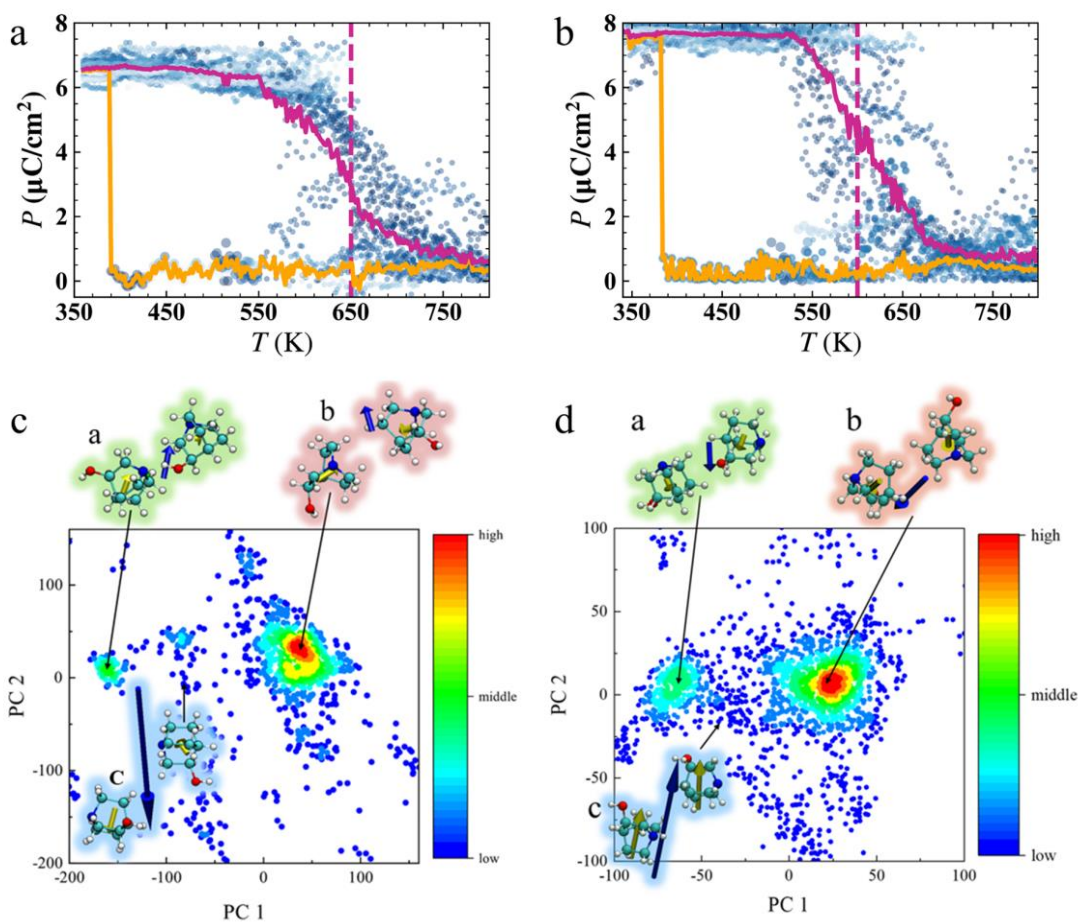

**Figure 5:** （a and b）: Phase transition behaviors of (*R*)/(*S*)-3-quinuclidinol. The yellow line shows the first phase transitions at 400 K of (*R*)-3-quinuclidinol and 398 K of (*S*)-3-quinuclidinol, which are highly consistent with the experimental values, and the purple line shows the other phase transition at a high temperature of about 600 K. (c and d): PCA for (*R*)- and (*S*)- isomers. The color bar shows the density of the points representing structures, and arrowed points (a) and (b) represent the characteristic structures for two paraelectric phases, and (c) represents the characteristic structure for the ferroelectric phase.

Figure 5a, b exhibits the spontaneous polarization of the (*R*)/(*S*)-3-quinuclidinol crystal with respect to temperature from REMD. Surprisingly, they both show two different critical points, 380 K (*R*)-/385 K (*S*)- and 650 K (*R*)-/600 K (*S*)-. The first one of each isomer is definitely the $T_c$ found from experimental work[14]. The second one is a novel finding since it has not been reported by experiments yet. Howeversince the decomposition temperatures for the crystals of both (*R*)- and (*S*)-

isomers are about 495 K, it would be a challenge for experimentalists to validate the second critical point. For validating our finding, we also performed cluster analysis forPCs, as shown in Figure 5c, d. Figures S7 and S8 display the proportion of the PCs in the variance of the original distribution. [39] The scattered points represent the structures from all of the REMD calculations described by PCs, and the color is produced by kernel density estimation. From the plots, it can be recognized that there are three major clusters for each isomer, denoted by a, b, and c. We display the characteristic structure of each cluster by picking the center of the highest density in each cluster in Figure 5c, d. The cluster c for each isomer is in the beginning ferroelectric phase, and both a and c are in the paraelectric phase. Therefore, it is clear that there are two different paraelectric phases corresponding to both $T_c$.

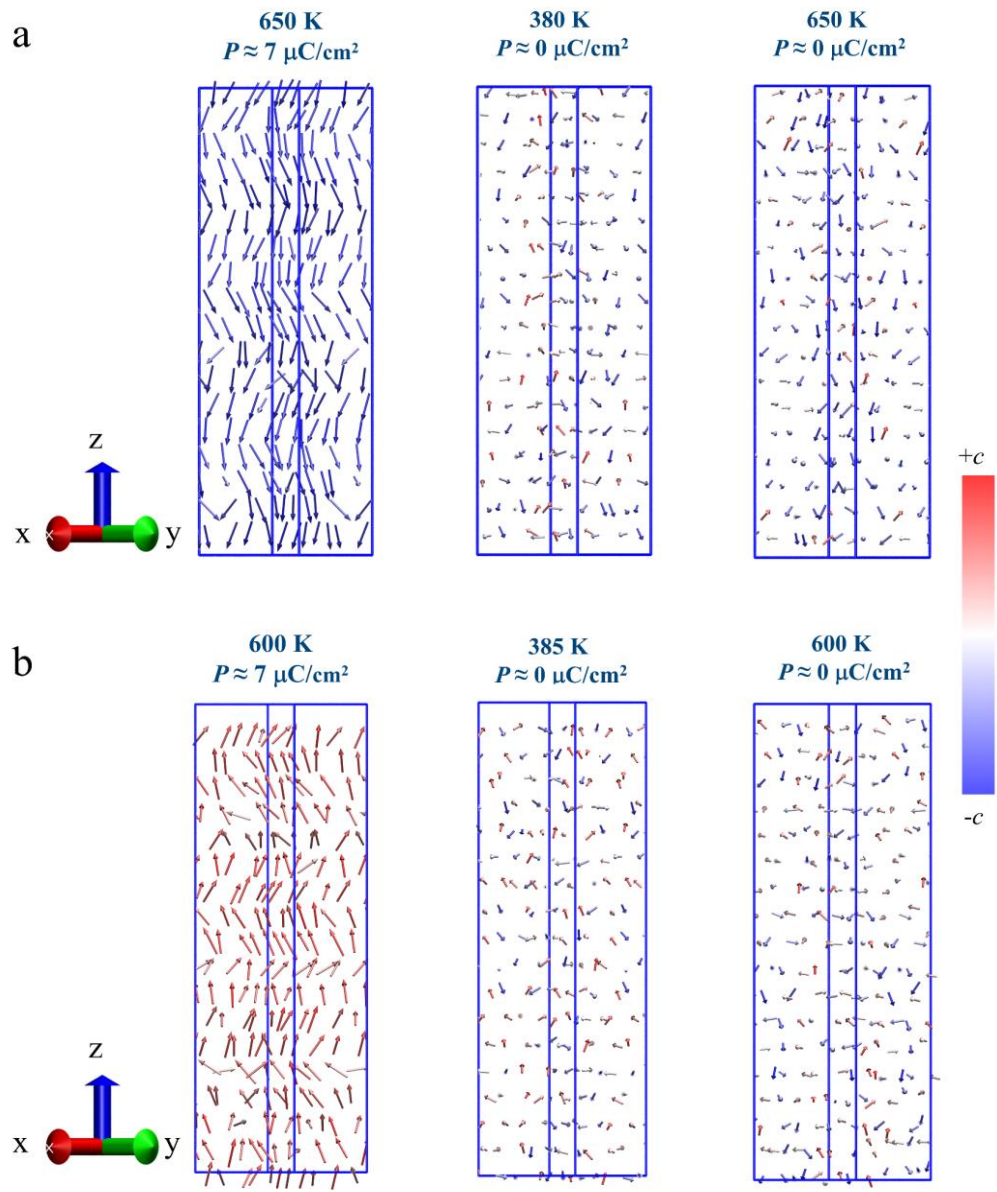

**Figure 6:** Dipole moments of each molecule in the three conformations of the $(R)/(S)$-3-quinuclidinol crystal are shown by arrows. (a): $(R)$-3-quinuclidinol. (b): $(S)$-3-quinuclidinol. From left to right: the ferroelectric phase, the paraelectric phase (a), and the paraelectric phase (b). Blue represents the negative direction along the $c$-axis, and red represents the positive direction along the $c$-axis.

For differentiating each phase, we also depicted dipole moments of six molecules within prime cells for all the three characteristic conformations.[5] As shown in Figure 6, the structures and corresponding dipole moments are shown alongside the clusters. The blue arrow represents the dipole moment of a single cell, and the yellow one represents the dipole moment of each molecule. All the three characteristic conformations are analyzed by Multiwfn.[50] The structures and corresponding dipole moments are shown alongside the clusters. Details are shown in Figures S5 and S6 and Tables S3 and S4. From these plots, we can understand that for both the paraelectric phases, the dipole moment within each prime cell is non-zero but randomly arranged, resulting in a total paraelectric phase. However, the difference between both of the paraelectric phases is hard to distinguish by eyes. By using the results of REMD, we also calculated the potential of mean force (PMF) of the phase transition using the multiple Bennett's acceptance ratio (MBAR) method.[51, 52] The free energy barrier for the (*S*)- enantiomer is 6.7 kcal/mol, slightly higher than that of the (*R*)-, 6.0 kcal/mol. The profile is shown in Figure 7. Since the difference between both barriers is small, we contribute the difficulty of phase transition of the (*S*)-enantiomer to the broad width of PMF. However, a small increase in its height hinders the system much to overcome it.

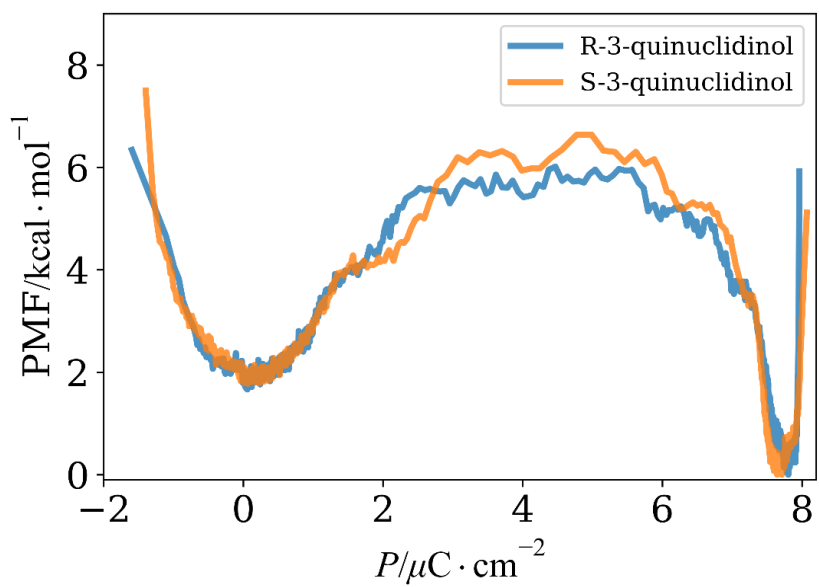

**Figure 7:** PMFobtained from MBAR calculations using REMD results, at the first $T_c$, 380 K.

## Conclusion

We investigated the origin of spontaneous polarization, the mechanism of both ferroelectric switching and phase transition for (*R*)/(*S*)-3-quinuclidinol, based on MD simulations with PCC. Our results validate the $T_c$ and $E_c$ for the crystals of both isomers, as high as 380 K (*R*)-/385 K (*S*)-, which is much higher than other chiral ferroelectrics and comparable to the classic inorganic ferroelectric barium titanate (393 K). In the study of $T_c$, we also find a novel critical point as high as about 600 K, higher than the decomposition temperature of (*R*)/(*S*)-3-quinuclidinol. We also find that the switching process is not according to the KAI model; hoping future work can provide a new model to explain it. In addition, the saturated polarization strength (up to 8 μC·cm$^{-2}$) is equivalent to the organic polymer ferroelectric PVDF[55], and the low coercive field (15 kV·cm$^{-1}$) can ensure that the ferroelectric polarization is easy to switch. This result reveals that (*R*)/(*S*)-3-quinuclidinol has good ferroelectricity, which is also necessary for ferroelectric devices.

**Competing interests**

The authors declare no competing interests.

**Corresponding author**

Correspondence to Yongle Li.

**Acknowledgments**

This study was funded by the National Natural Science Foundation of China (No. 21503130 and 11674212). We appreciate the Ziqiang 4000 of Shanghai University for high-performance computing services. We thank Ye Mei (East China Normal University) and Ya Gao (Shanghai University of Engineering Science) for some helpful discussions.

**Supporting Information:** The detailed crystal structures and the density of states of (*R*)/(*S*)-3-quinuclidinol crystal and the simulation results of dynamic simulation.

Supporting Information for

# **Mechanisms of Molecular Ferroelectrics Made Simple**

*Xiaoqing Zhu,[ab] Wenbin Fan,[ab] Wei Ren,[ab] and Yongle Li*[ab]*

[a] Department of Physics, Shanghai University, Shanghai 200444, China.
[b] Department of Physics, International Center of Quantum and Molecular Structures, Shanghai University, Shanghai 200444, China.
*Email: yongleli@shu.edu.cn;

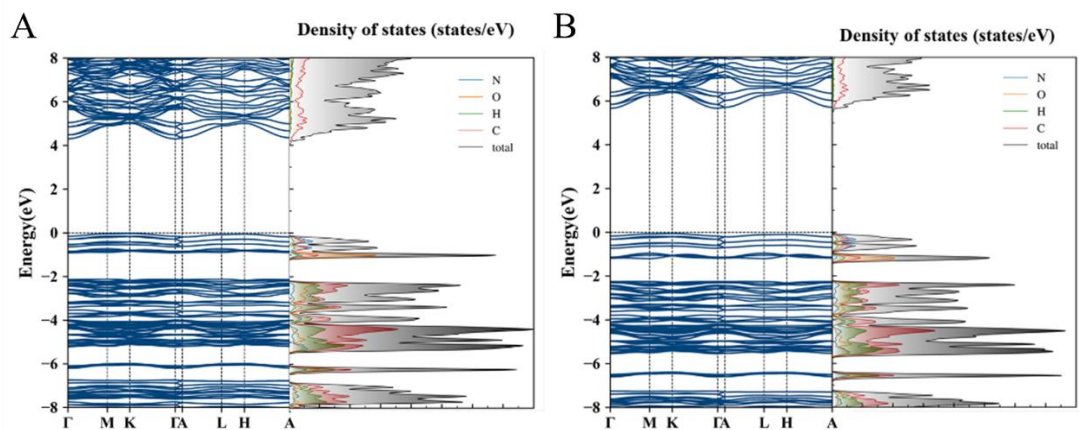

**Figure S1:** The density of states (DOS) of *R*-3-quinuclidinol calculated through (A) PBE exchange-correlation functionals and (B) Heyd-Scuseria-Ernzerhof functionals. The contributions from nitrogen, oxygen, hydrogen, and carbon atoms are highlighted by blue, orange, green, and purple lines, respectively.

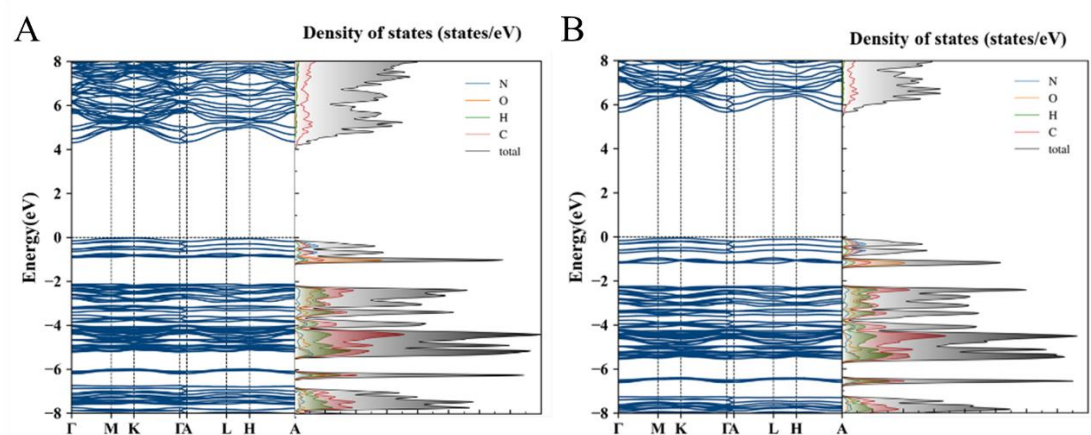

**Figure S2:** The density of states (DOS) of *S*-3-quinuclidinol as calculated through (A) PBE exchange-correlation functionals and (B) Heyd-Scuseria-Ernzerhof functionals. The contributions from nitrogen, oxygen, hydrogen, and carbon atoms are highlighted by blue, orange, green, and purple lines, respectively.

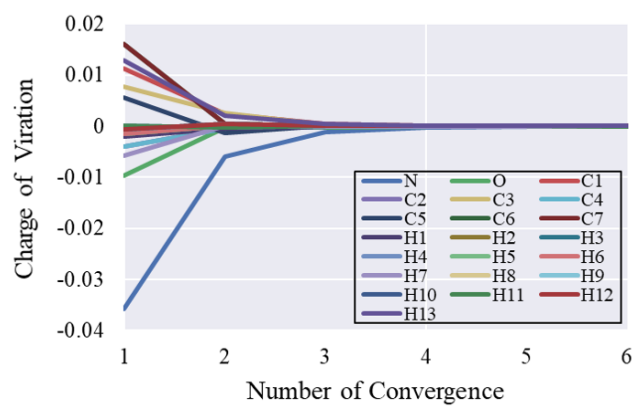

**Figure S3**: The PCC charge convergence of (*R*)-3-quinuclidinol crystal.

**Table S1**. The charge convergence of (*R*)-3-quinuclidinol crystal

| number Atomic named | 1 | 2 | 3 | 4 | 5 | 6 | 7 | 8 |
|---|---|---|---|---|---|---|---|---|
| O1 | -0.790 | -0.973 | -0.984 | -0.985 | -0.985 | -0.986 | -0.986 | -0.986 |
| N1 | -0.664 | -0.817 | -0.844 | -0.848 | -0.849 | -0.849 | -0.849 | -0.849 |
| C1 | 0.313 | 0.278 | 0.284 | 0.285 | 0.285 | 0.285 | 0.285 | 0.285 |
| C2 | 0.011 | -0.023 | -0.032 | -0.032 | -0.032 | -0.032 | -0.032 | -0.032 |
| C3 | 0.335 | 0.476 | 0.486 | 0.488 | 0.489 | 0.489 | 0.489 | 0.489 |
| H5 | -0.044 | -0.058 | -0.059 | -0.060 | -0.060 | -0.060 | -0.060 | -0.060 |
| C4 | -0.021 | 0.004 | 0.007 | 0.008 | 0.008 | 0.008 | 0.008 | 0.008 |
| C5 | -0.021 | 0.004 | 0.007 | 0.008 | 0.008 | 0.008 | 0.008 | 0.008 |
| C6 | 0.119 | 0.137 | 0.145 | 0.146 | 0.146 | 0.146 | 0.146 | 0.146 |
| C7 | 0.119 | 0.137 | 0.145 | 0.146 | 0.146 | 0.146 | 0.146 | 0.146 |

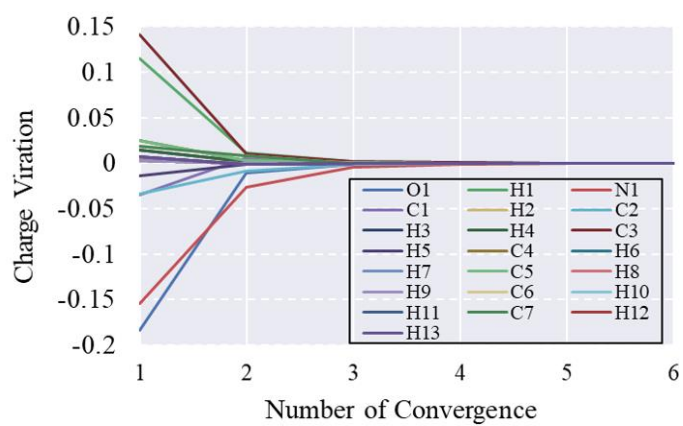

**Figure S4:** The PCC charge convergence of (*S*)-3-quinuclidinol.

**Table S2.** The charge convergence of (*S*)-3-quinuclidinol crystal

| Atomic named \ number | 1 | 2 | 3 | 4 | 5 | 6 | 7 | 8 |
|---|---|---|---|---|---|---|---|---|
| O1 | -0.790 | -0.973 | -0.984 | -0.985 | -0.985 | -0.986 | -0.986 | -0.986 |
| N1 | -0.664 | -0.817 | -0.844 | -0.848 | -0.849 | -0.849 | -0.849 | -0.849 |
| C1 | 0.313 | 0.278 | 0.284 | 0.285 | 0.285 | 0.285 | 0.285 | 0.285 |
| C2 | 0.011 | -0.023 | -0.032 | -0.032 | -0.032 | -0.032 | -0.032 | -0.032 |
| C3 | 0.335 | 0.476 | 0.486 | 0.488 | 0.489 | 0.489 | 0.489 | 0.489 |
| H5 | -0.044 | -0.058 | -0.059 | -0.060 | -0.060 | -0.060 | -0.060 | -0.060 |
| C4 | -0.021 | 0.004 | 0.007 | 0.008 | 0.008 | 0.008 | 0.008 | 0.008 |
| C5 | -0.021 | 0.004 | 0.007 | 0.008 | 0.008 | 0.008 | 0.008 | 0.008 |
| C6 | 0.119 | 0.137 | 0.145 | 0.146 | 0.146 | 0.146 | 0.146 | 0.146 |
| C7 | 0.119 | 0.137 | 0.145 | 0.146 | 0.146 | 0.146 | 0.146 | 0.146 |

**Table S3.** Acceptance ratios of replica corresponding to pairs of neighboring temperatures for (*R*)-3-quinuclidinol crystal.

| Pair of temperatures | Acceptance ratio | Pair of temperatures | Acceptance ratio |
|---|---|---|---|
| 350.00 K↔358.18 K | 0.263 | 544.59 K↔556.28 K | 0.263 |
| 358.18 K↔366.50 K | 0.239 | 556.28 K↔568.18 K | 0.248 |
| 366.50 K↔374.97 K | 0.251 | 568.18 K↔580.29 K | 0.261 |
| 374.97 K↔383.59 K | 0.264 | 580.29 K↔592.62 K | 0.257 |
| 383.59 K↔392.37 K | 0.274 | 592.62 K↔605.18 K | 0.276 |
| 392.37 K↔401.31 K | 0.278 | 605.18 K↔617.96 K | 0.273 |
| 401.31 K↔410.42 K | 0.275 | 617.96 K↔630.99 K | 0.270 |
| 410.42 K↔419.68 K | 0.273 | 630.99 K↔644.24 K | 0.267 |
| 419.68 K↔429.10 K | 0.239 | 644.24 K↔657.74 K | 0.261 |
| 429.10 K↔438.71 K | 0.274 | 657.74 K↔671.47 K | 0.258 |
| 438.71 K↔448.48 K | 0.255 | 671.47 K↔685.46 K | 0.272 |
| 448.48 K↔458.42 K | 0.254 | 685.46 K↔699.71 K | 0.251 |
| 458.42 K↔468.48 K | 0.275 | 699.71 K↔714.21 K | 0.266 |
| 468.48 K↔478.79 K | 0.270 | 714.21 K↔728.98 K | 0.265 |
| 478.79 K↔489.28 K | 0.272 | 728.98 K↔744.01 K | 0.284 |
| 489.28 K↔499.96 K | 0.269 | 744.01 K↔759.36 K | 0.280 |
| 499.96 K↔510.78 K | 0.262 | 759.36 K↔774.94 K | 0.273 |
| 510.78 K↔521.85 K | 0.277 | 774.94 K↔790.87 K | 0.279 |
| 521.85 K↔533.12 K | 0.271 | 790.87 K↔799.04 K | 0.250 |
| 533.12 K↔544.59 K | 0.246 | 799.04 K↔800.00 K | 0.270 |

**Table S4.** Acceptance ratios of replicas corresponding to pairs of neighboring temperatures for (*S*)-3-quinuclidinol crystal.

| Pair of temperatures | Acceptance ratio | Pair of temperatures | Acceptance ratio |
|---|---|---|---|
| 350.00 K ↔ 358.18 K | 0.270 | 544.59 K ↔ 556.28 K | 0.251 |
| 358.18 K ↔ 366.50 K | 0.255 | 556.28 K ↔ 568.18 K | 0.248 |
| 366.50 K ↔ 374.97 K | 0.266 | 568.18 K ↔ 580.29 K | 0.266 |
| 374.97 K ↔ 383.59 K | 0.263 | 580.29 K ↔ 592.62 K | 0.271 |
| 383.59 K ↔ 392.37 K | 0.271 | 592.62 K ↔ 605.18 K | 0.253 |
| 392.37 K ↔ 401.31 K | 0.253 | 605.18 K ↔ 617.96 K | 0.251 |
| 401.31 K ↔ 410.42 K | 0.259 | 617.96 K ↔ 630.99 K | 0.253 |
| 410.42 K ↔ 419.68 K | 0.261 | 630.99 K ↔ 644.24 K | 0.252 |
| 419.68 K ↔ 429.10 K | 0.257 | 644.24 K ↔ 657.74 K | 0.251 |
| 429.10 K ↔ 438.71 K | 0.253 | 657.74 K ↔ 671.47 K | 0.266 |
| 438.71 K ↔ 448.48 K | 0.252 | 671.47 K ↔ 685.46 K | 0.262 |
| 448.48 K ↔ 458.42 K | 0.251 | 685.46 K ↔ 699.71 K | 0.235 |
| 458.42 K ↔ 468.48 K | 0.256 | 699.71 K ↔ 714.21 K | 0.250 |
| 468.48 K ↔ 478.79 K | 0.257 | 714.21 K ↔ 728.98 K | 0.243 |
| 478.79 K ↔ 489.28 K | 0.260 | 728.98 K ↔ 744.01 K | 0.263 |
| 489.28 K ↔ 499.96 K | 0.250 | 744.01 K ↔ 759.36 K | 0.232 |
| 499.96 K ↔ 510.78 K | 0.262 | 759.36 K ↔ 774.94 K | 0.214 |
| 510.78 K ↔ 521.85 K | 0.259 | 774.94 K ↔ 790.87 K | 0.261 |
| 521.85 K ↔ 533.12 K | 0.263 | 790.87 K ↔ 799.04 K | 0.251 |
| 533.12 K ↔ 544.59 K | 0.246 | 799.04 K ↔ 800.00 K | 0.231 |

**Table S5.** The electric field strength (*E*) and the characteristic switching time ($t_0$) of (*R*)/(*S*)-3-quinuclidinol.

| $E$ / V·nm$^{-1}$ | (*R*)- $t_0$ / ns | (*S*)- $t_0$ / ns | (*R*)- n | (*S*)- n |
|---|---|---|---|---|
| 17 | 6.02 | 7.23 | 20.33 | 20.03 |
| 18 | 3.24 | 3.05 | 20.01 | 19.56 |
| 19 | 1.71 | 1.07 | 19.60 | 19.30 |
| 20 | 0.95 | 0.30 | 14.47 | 14.38 |
| 21 | 0.47 | 0.145 | 12.38 | 9.08 |
| 22 | 0.28 | 0.065 | 6.05 | 8.03 |

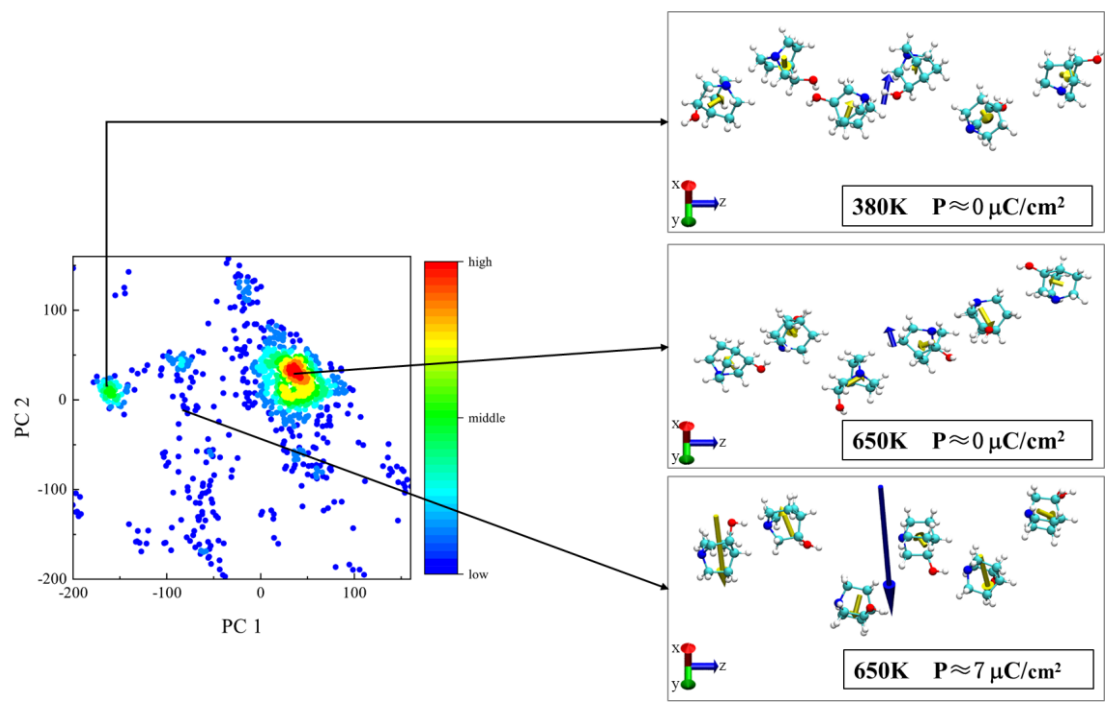

**Figure S5:** The dipole moments of three single molecule of (*R*)-3-quinuclidinol crystal are shown by yellow arrows, and the total dipole moments of the (*R*)-3-quinuclidinol crystal are shown by blue arrows.

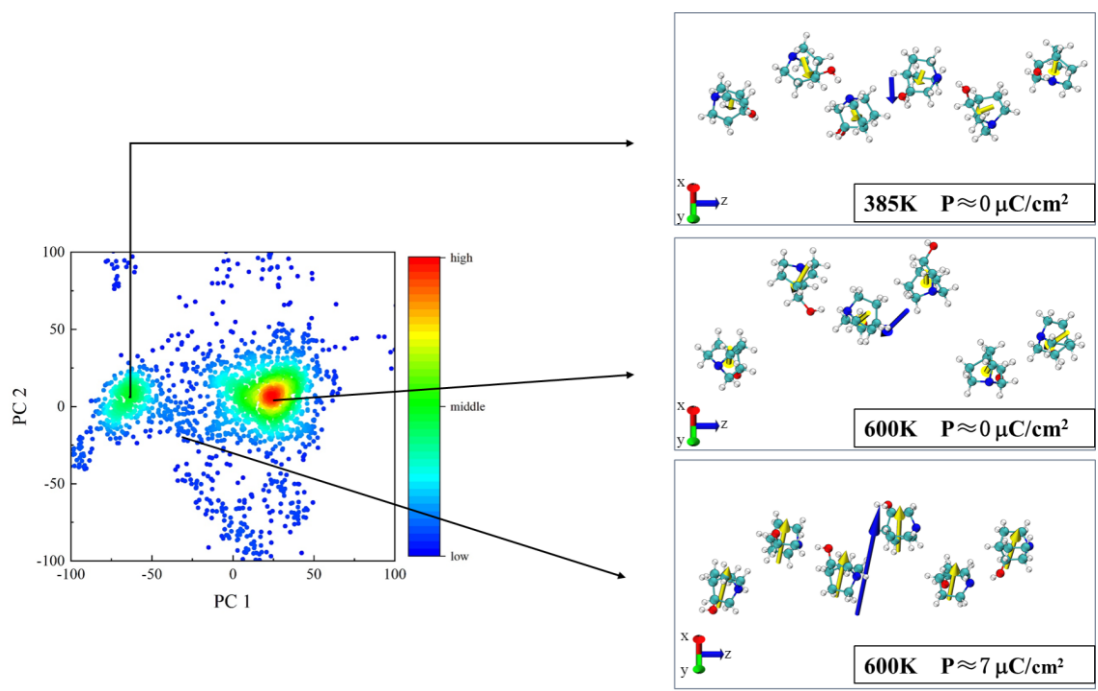

**Figure S6:** The dipole moments of each monomer molecule in the three conformations of (*S*)-3-quinuclidinol crystal are shown by yellow arrows, and the total dipole moments of the (*S*)-3-quinuclidinol crystal are shown by blue arrows.

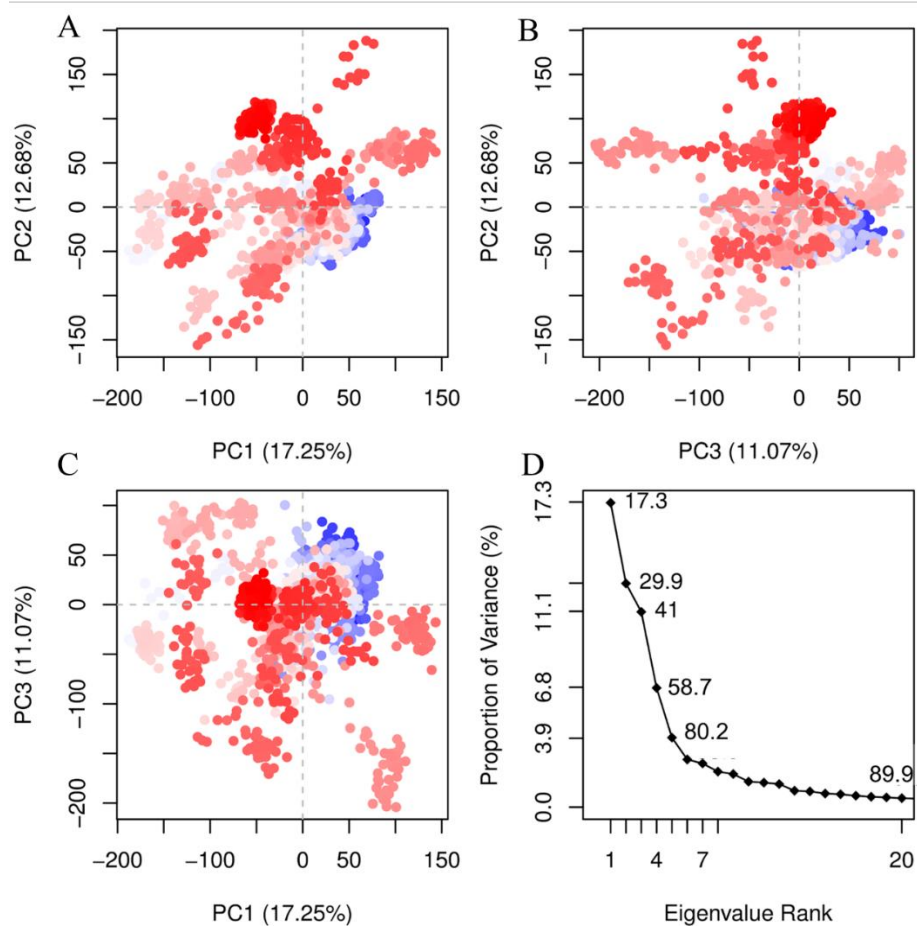

**Figure S7:** For replica-exchange dynamics simulation of (*R*)-3-quinuclidinol crystal, A is the distribution plot for the 1st and the 2nd PCs. B is the distribution plot for the 2nd and 3rd PCs. C is the distribution for the 1st and 3rd PCs. The D shows the first five PCs can cover over 80% kinetic information.

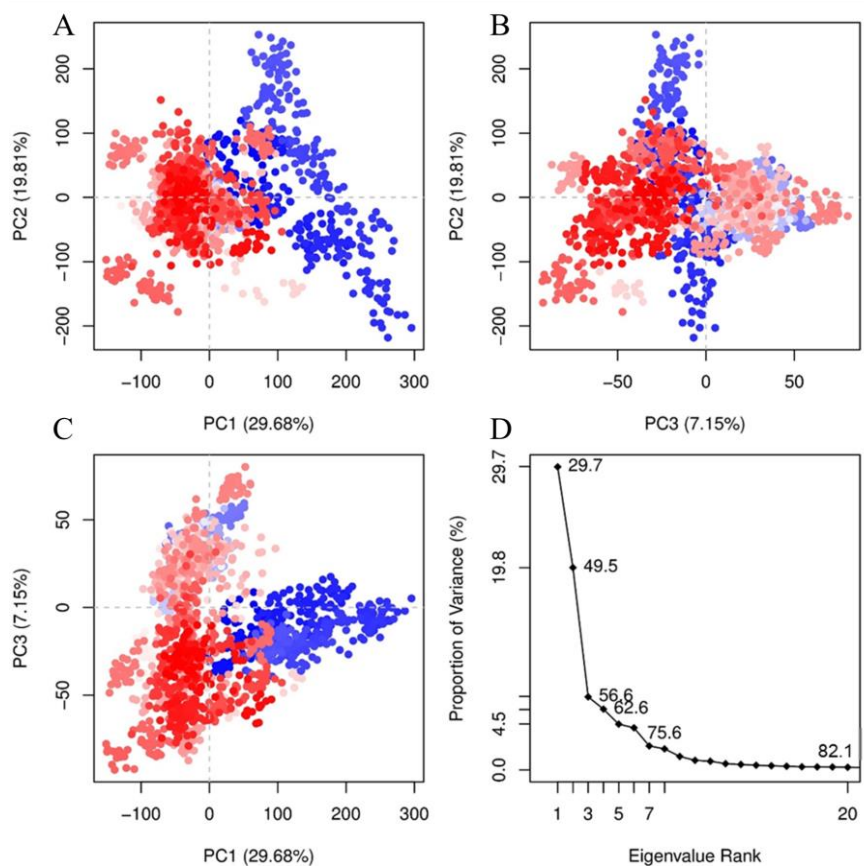

**Figure S8:**   For replica-exchange dynamics simulation of (*S*)-3-quinuclidinol crystal, A is the distribution plot for the 1st and the 2nd PCs. B is the distribution plot for the 2nd and 3rd PCs. C is the distribution for the 1st and 3rd PCs. The D shows the first five PCs can cover over 70% kinetic information.